\documentclass[twocolumn]{aastex631}

\usepackage{amssymb,amsmath}

\shorttitle{AASTeX v6.3.1 Sample article}
\shortauthors{Zhang et al.}

\graphicspath{{./}{figures/}}

\begin{document}

\title{Quasi-periodic oscillations of flare loops and slipping motion of ribbon substructures during a C-class flare}

\author[0000-0001-5933-5794]{Yining Zhang}
\affiliation{National Astronomical Observatories, Chinese Academy of Sciences,
Beijing 100101, People's Public of China}
\affiliation{School of Astronomy and Space Science,
University of Chinese Academy of Sciences,
Beijing 100049, People's Public of China}

\correspondingauthor{Ting Li}
\email{liting@nao.cas.cn}
\correspondingauthor{Jing Ye}
\email{yj@ynao.ac.cn}

\author[0000-0001-6655-1743]{Ting Li}
\affiliation{National Astronomical Observatories, Chinese Academy of Sciences,
Beijing 100101, People's Public of China}
\affiliation{School of Astronomy and Space Science,
University of Chinese Academy of Sciences,
Beijing 100049, People's Public of China}
\affiliation{State Key Laboratory of Solar Activity and Space Weather, Beijing 100190,
People's Republic of China}

\author[0000-0002-5983-104X]{Jing Ye}
\affiliation{Yunnan Observatories, Chinese Academy of Sciences, Kunming, Yunnan 650216, People's Republic of China}
\affiliation{Yunnan Key Laboratory of Solar Physics and Space Sciences, P.O. Box 110, Kunming, Yunnan 650216, People's Republic of China}
\affiliation{Yunnan Province China-Malaysia HF-VHF Advanced Radio Astronomy Technology International Joint Laboratory, Kunming, Yunnan 650216, People's Republic of China}

\begin{abstract}
Quasi-periodic oscillations in solar flaring emission
have been observed over the past few decades. To date, the underpinning processes resulting in the quasi-periodic oscillations remain unknown. In this paper, we report a unique event that exhibits both the long-duration quasi-periodic intensity oscillations of flare loops and the quasi-periodic slipping motion of ribbon substructures during a C9.1-class flare (SOL2015-03-15-T01:15), using the observations from Solar Dynamics Observatory and Interface Region Imaging Spectrograph. The high-temperature flare loops rooted in the straight part of ribbons display a ``bright-dim" intensity oscillation, with a period of about 4.5 minutes. The oscillation starts just after the flare onset and lasts over 3 hours. Meanwhile, the substructures within the ribbon tip display the quasi-periodic slipping motion along the ribbon at 1400 \AA~images which has a similar periodicity to the stationary intensity oscillation of the flare loops in the straight part of the flare ribbons. We suggest that the quasi-periodic pattern is probably related to the loop-top dynamics caused by the reconnection outflow impinging on the flare loops.

\end{abstract}


\section{Introduction} \label{sec:intro}
Solar flares are the most energetic events in the solar system. They can release up to $10^{33}$ ergs of energy \citep{Emslie2012} into the interplanetary space, which can have a significant impact on the space environment and human activities. However, the exact mechanism of releasing such enormous energy in such a short period by solar flares remains elusive. It is widely believed that the magnetic reconnection process plays a crucial role in converting the magnetic energy into other forms of energies, such as kinetic and thermal energies. Two-dimensional (2D) magnetic reconnection has been incorporated into the classical CSHKP model of solar flare \citep{Carmichael1964,Sturrock1966,Hirayama1974,Kopp1976}. Previous studies have shown many observational evidences for the 2D CSHKP model, including features such as the cusp structure detected at the flare loop top \citep{Tan2020} and the discernible presence of a large-scale current sheet \citep{Ciaravella2008,Cheng2018}.

However, magnetic reconnection is intrinsically a three-dimensional (3D) process \citep{Priest1995,Janvier2013}. Many observational characteristics cannot be simply explained by a 2D model, including the formation of magnetic flux ropes and the J-shaped flare ribbons \citep{Li2021}. Therefore, it is necessary to develop a 3D reconnection model of the solar flare. \cite{Aulanier2012} and \citet{Janvier2013} presented zero-beta magnetohydrodynamic (MHD) numerical simulations to study the evolution of the magnetic field lines. Their results ascertain that the 3D magnetic reconnection process produces the slipping motion of the magnetic field lines. In 2D reconnection models, the reconnection process mainly happens near the magnetic null-point, where the change of the connectivity of magnetic field is not continuous. In 3D, magnetic reconnection can also occur along quasi-separatrix layers (QSLs) \citep{Priest1995,Demoulin1996} where the magnetic connectivity is continuous but with a strong gradient. Magnetic field lines
undergo a continuous series of reconnections as they cross QSLs. The continuous restructuring of field lines along the QSLs results in the apparent slipping motion of field line footpoints. Thus the slipping motion is regarded as a key feature for 3D magnetic reconnection different from 2D theories \citep{Aulanier2005,Aulanier2006}. Recently, with the improvement of observational instruments, the slipping motion has been verified by some previous works. \cite{Aulanier2007} first reported the slipping motion of coronal loops observed by the X-Ray Telescope (XRT) onboard Hinode. \citet{Dudik2014} studied an X-class flare and found that the flare loops underwent the slipping motion along the ribbons with the velocity of several tens of km s$^{-1}$ by Solar Dynamics Observatory (SDO, \citealt{Pesnell2012})/Atmospheric Imaging Assembly (AIA, \citealt{Lemen2012}). Furthermore, \cite{Li2015} analyzed an X-class flare and found the quasi-periodic slipping motion of the flare loops along the flare ribbons by using data from SDO and Interface Region Imaging Spectrograph (IRIS, \citealt{Depontieu2014}) together, which shows the period of 3-6 minutes. 

Quasi-periodic pulsations (QPPs) in solar flaring emission
have been observed over the past several decades. One interesting topic is to understand the origin of QPPs observed at different locations (\citealt{McLaughlin2018,Zimovets2021}), such as the above-the-loop-top(ALT) region and the flare ribbons. The QPPs at the ALT have been identified with periods ranging from 74s to 400s in many previous studies (\citealt{Kupriyanova2019,Cai2019,Reeves2020}). The supersonic reconnection outflows from the current sheet to the flare loops generate termination shocks and excite local oscillation in the ALT region \citep{Takasao2015,Shen2018,Zhang2022}. The footpoints of flare loops also show periodic oscillations especially at microwave and hard X-Ray wave bands. \citet{Huang2016} detected QPPs at flare loop footpoints and loop top during an M7.7 class flare with a period of 270 s in X-Ray and microwave wavebands. It is mostly caused by the quasi-periodic process of acceleration or transportation of energetic electrons from the looptop to the footpoints of the loops. \citet{Brannon2015} and \citet{Parker2017} investigated the quasi-periodic sawtooth structures in the flare ribbons and suggested that they are caused by Kelvin-Helmholtz or tearing mode instabilities. In a word, previous observation of QPPs focused on the loop top or loop footpoints. As for the whole flare loops, \citet{Tian2016} detected the oscillations of flare loops with a 25s period in both the intensity and Doppler shift of the Fe XXI 1354.08 \AA ~line during an M1.6 class flare. They strongly support the interpretation of QPPs as global fast sausage oscillations. To our knowledge, direct observations of intensity oscillations of the entire flare loop are relatively rare.

Here, we report a unique event that displays the quasi-periodic patterns in both the intensity oscillations of the flare loops and the slipping motion of the flare ribbons. This paper is organized as follows: In Section \ref{sec:obsresults} we describe the observational results and the related analysis of the periodicity of this event. In Section \ref{sec:discussion}, we discuss the potential mechanisms responsible for this long-duration quasi-periodic behavior. Finally, Section \ref{sec:summary} gives the summary and conclusion of this work.

\section{Observational Results and Analysis}  \label{sec:obsresults}

\subsection{Overview of the C9.1-class flare}
\begin{figure*}[!ht]
    \centering
    \includegraphics[width=0.85\textwidth,trim=2 50 5 50,clip]{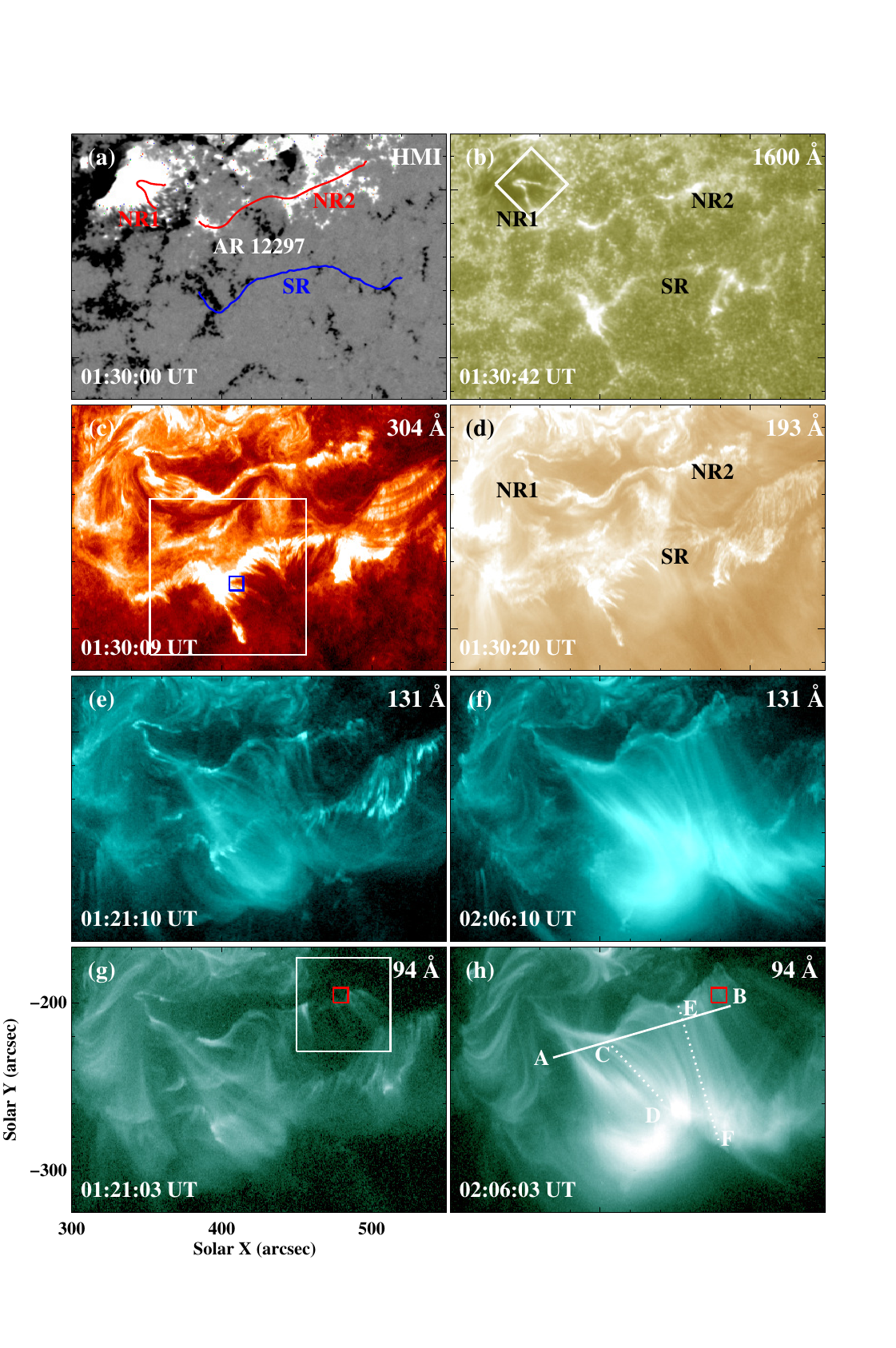}
    \caption{Multi-wavelength images of flare ribbons and flare loops of the C9.1-class flare on 2015 March 15. Panel (a): SDO/HMI line-of-sight magnetogram with the field of view (FOV) of ~240" $\times$ 170". Red and blue lines denote the flare ribbons NR1, NR2 and SR. Panels (b)-(h): SDO/AIA 1600\AA, 304\AA, 193\AA, 131\AA ~and 94\AA ~images with the same FOV as panel (a). White rectangle in panel (b) denotes the FOV of Figures \ref{fig:fig8}(b)-(h). White rectangle and blue square in panel (c) denote the FOV of Figures \ref{fig:fig6}(a)-(i) and the same blue square in Figures \ref{fig:fig6}(g)-(i), respectively. White rectangle and red squares in panels (g)-(h) denote the FOV of Figures \ref{fig:fig4}(a)-(h) and the same red square in Figures \ref{fig:fig4}(a)-(h), respectively. The straight white solid lines indicate the slit `A-B', which is used to obtain the stack plots in Figure \ref{fig:fig2}. The white dotted lines in panel (h) indicate the slits `C-D' and `E-F', which are used to obtain the stack plots in Figures \ref{fig:fig4}(a)-(b) and (c)-(d), respectively. The animation of this figure includes HMI magnetogram, AIA 94, 131, 193, 304 and 1600 \AA ~images from 01:00 UT to 03:00 UT with a video duration of 8s.}
    \label{fig:fig1}
\end{figure*}

On 2015 March 15, a C9.1-class solar flare occurred within the domain of NOAA Active Region (AR) 12297 (see Figure \ref{fig:fig1}). As discerned from GOES soft X-Ray flux data, the C9.1-class flare initiated at 01:15UT and reached its peak at around 02:13UT (indicated by the orange line in Figure \ref{fig:fig4}(d)). As shown in Figure \ref{fig:fig1}(a) of Helioseismic and Magnetic Imager (HMI) magnetogram, the flare ribbons are composed of two quasi-parallel ribbons of NR2 and SR, and also a hook-shaped ribbon NR1 locating at the positive-polarity sunspot. The flare loops connecting NR2 and SR show the intensity oscillations as observed in AIA 94/131\AA~wavelengths.

\subsection{Stationary long-duration quasi-periodic intensity oscillation of flare loops}
\begin{figure*}[!ht]
    \centering
    \includegraphics[width=0.90\textwidth,trim=5 5 5 10,clip]{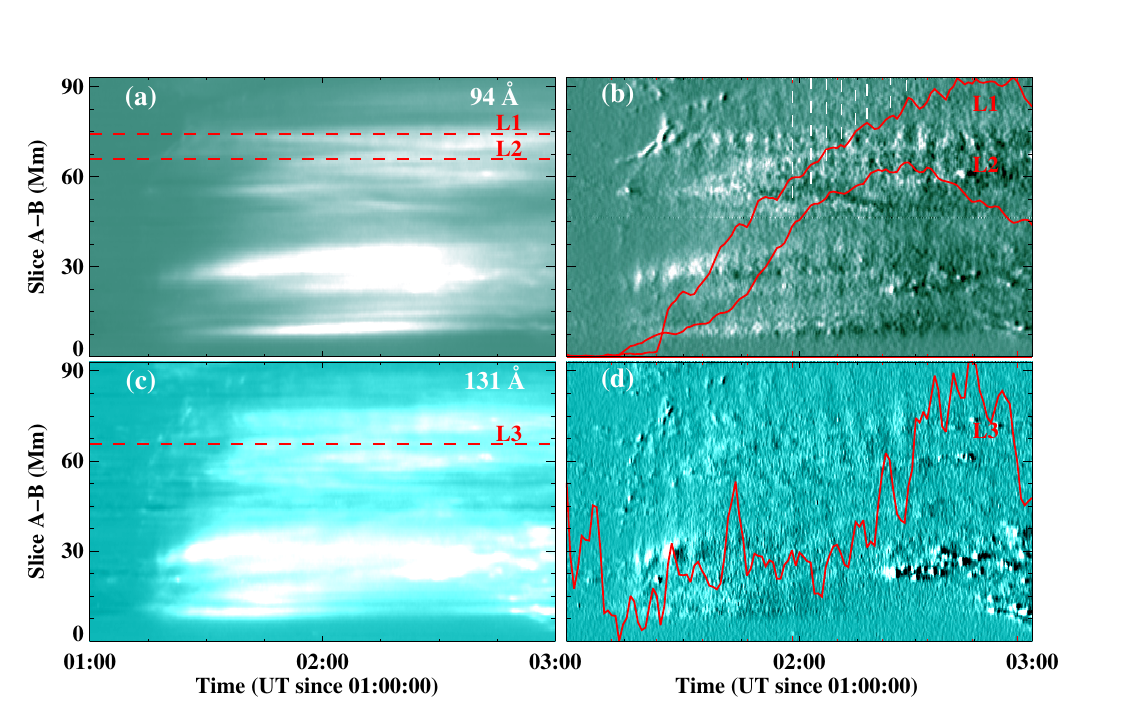}
    \caption{Panels (a)-(b): Stack plots along slice `A-B' at AIA 94 \AA ~of base difference and running difference, respectively. Panels (c)-(d): Stack plots along slice `A-B' at AIA 131 \AA ~of base difference and running difference, respectively. Red dashed lines in panel (a) denote the positions to calculate the light curves L1 and L2 in panel (b). Similarly, the red dashed line in panel (c) denotes the position to obtain the light curve L3 in panel (d).}
    \label{fig:fig2}
\end{figure*}

The stationary intensity oscillations of the flare loops lasted for a quite long time (over 3 hours). We draw the stack plots along the flare loops with various directions (see the white solid and dotted lines in Figures \ref{fig:fig1}(h)). From the stack plots of slices `A-B' near ribbon NR2 (see Figure \ref{fig:fig2}), a group of stripes appears in 94\AA ~and 131\AA ~stack plots, indicating the quasi-periodic patterns in the high-temperature ([log $T$(K)$\sim$6.85] \& [log $T$(K)$\sim$7.05], respectively) flare loops \citep{O'Dwyer2010}. We also obtained the intensity variations along straight lines L1, L2 and L3 at the base difference stack plots (Figures \ref{fig:fig2}), which are plotted at the running difference stack plots (light curves L1-L3 in Figures \ref{fig:fig2} (b)-(d)). The intensity oscillation can be clearly discerned with the time interval between neighboring peaks of around 5 minutes. In addition, the quasi-periodic intensity oscillations also appear near SR. The long-duration stationary intensity oscillations near NR2 and SR are strong indications that the quasi-periodic behavior appears in the whole flare loop region.

\begin{figure*}[!ht]
    \centering
    \includegraphics[width=0.90\textwidth]{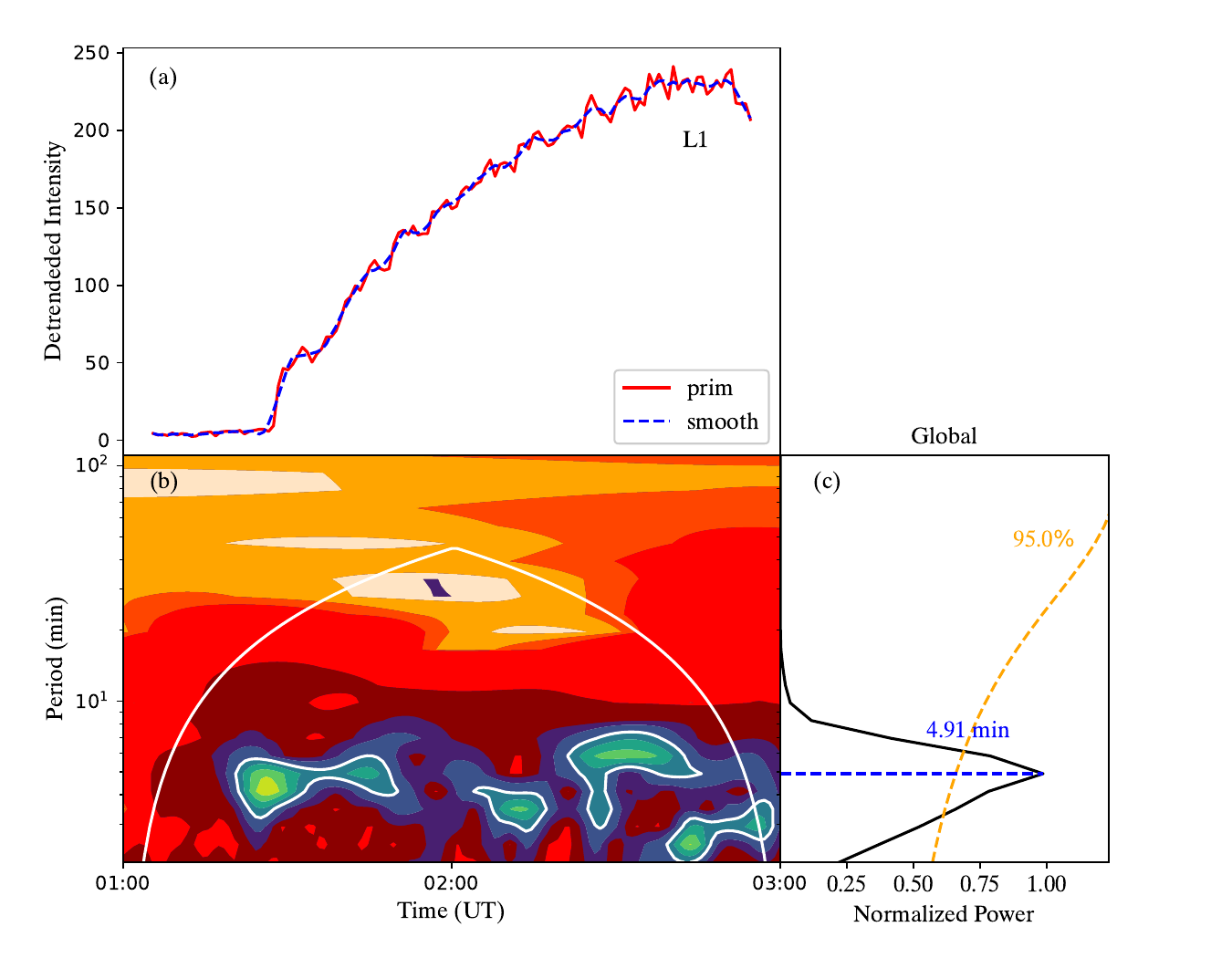}
    \caption{Panel (a): The normalized intensity of L1 (red solid line) shown in Figures \ref{fig:fig2}(a) and the smoothed light curve (blue dashed line), respectively. Panel (b): The wavelet power map of the detrended intensity which is calculated by the difference of the two light curves in panel (a) from 01:00 to 03:00 UT. Panel (c): The corresponding global power. The orange dashed line represents the 95\% confidence level, and the blue dashed line denotes the period over the 95\% confidence level with the max power (4.91 minutes).}
    \label{fig:fig3}
\end{figure*}

From Figures \ref{fig:fig2}(a)-(b), we can see that the normalized intensity along slice `A-B' shows an apparent oscillation in the light curves. To investigate the periodicity of the flare loops in SDO/AIA 94\AA ~during the C9.1 flare, we perform a wavelet analysis based on the Morlet transform via Python \citep{Torrence1998}. To highlight the short-period oscillations and to enhance analytical reliability, the detrended light curve, differentiating the blue dashed line from the red solid line in Figure \ref{fig:fig3}(a), was used for the wavelet analysis \citep{Kupriyanova2010, Kupriyanova2013, Gruber2011, Auchere2016,LiD2020}. It is evident that after about 01:15UT, the `L1' curve undergoes a rising stage with apparent oscillations in the light curve. The Morlet spectrum is given in Figure \ref{fig:fig3}(b) from 01:00 to 03:00UT, where the white line is the 95\% confidence level of the spectrum. As a result of the global power map computation, the blue dashed line signifies the epoch of maximal power, corresponding to an
oscillation period of approximately 4.9 minutes. It is noteworthy that its power surpasses the 95\% confidence level.

\begin{figure*}[!ht]
    \centering
    \includegraphics[width=0.90\textwidth,trim=5 10 5 20,clip]{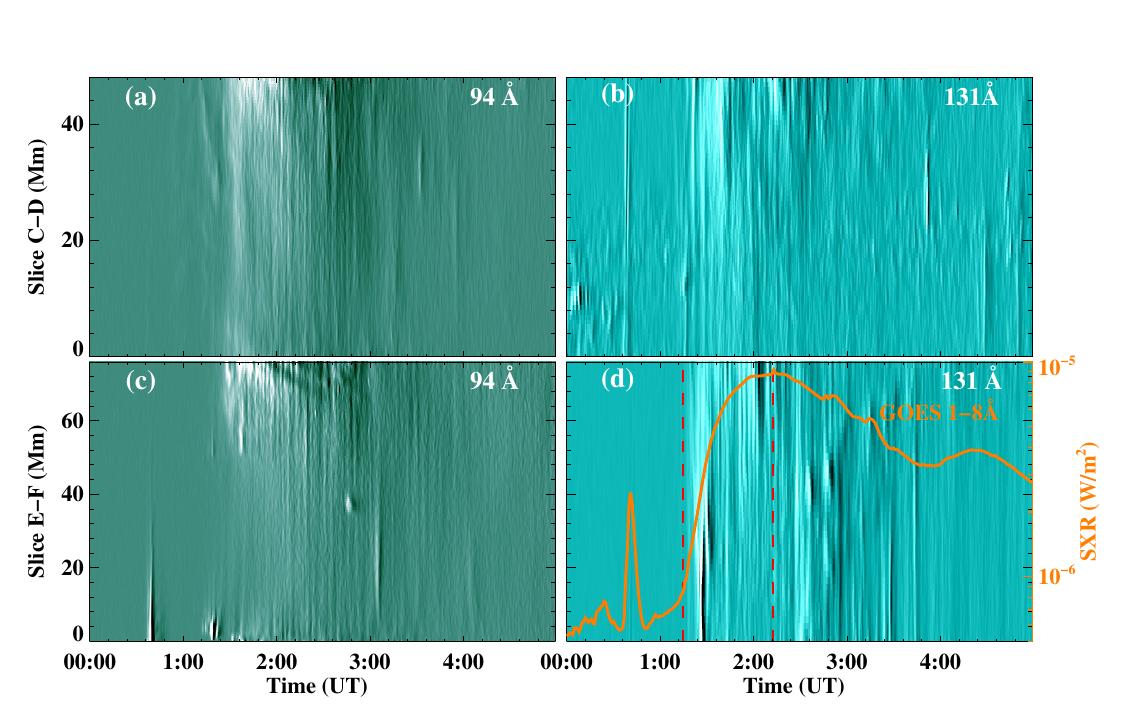}
        \caption{Panels (a)-(b): stack plots along slice `C-D' (white dotted line in Figure \ref{fig:fig1}(h)) of running difference images at 94\AA~ and 131\AA~, respectively. Panels (c)-(d): stack plots along slices `E-F' (white dotted line in Figure \ref{fig:fig1}(h)) of running difference images in 94\AA ~and 131\AA, respectively. Orange curve in panel (d) indicates the GOES SXR 1-8\AA ~flux. The red dashed lines in panel (d) denote the onset and the peak time of the flare, respectively.}
    \label{fig:fig4}
\end{figure*}

We also draw the stack plots (Figure \ref{fig:fig4}) along slices `C-D' and `E-F' along the flare loops (indicated by the white dotted lines in Figure \ref{fig:fig1}(h)). The bright-dim pattern lasted for a quite long duration from about 01:15 to 04:40UT in high-temperature wavebands, which is over 3 hours. The stationary oscillation of the flare loops almost covers the impulsive and decay phases of the flare (see Figure \ref{fig:fig4}(d)). To our knowledge, it is a persistent and rare flare event with this long duration pulsations.

\begin{figure*}[!ht]
    \centering
    \includegraphics[width=0.90\textwidth,trim=5 5 5 10,clip]{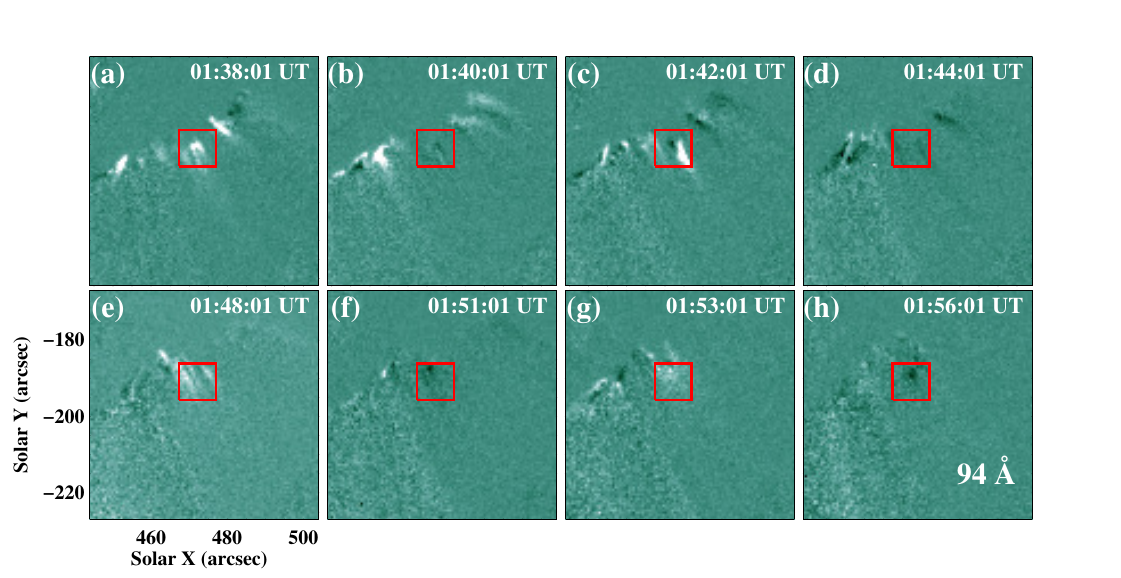}
    \caption{Running difference images at 94\AA ~with the FOV shown in Figure \ref{fig:fig1}(g). Red squares in panels (a)-(h) highlight the quasi-periodic intensity oscillation of the flare loop.}
    \label{fig:fig5}
\end{figure*}

In order to further investigate the temporal variations of the high-temperature flare loops, we make running-difference images of SDO/AIA 94\AA ~images (with the FOV in Figure \ref{fig:fig1}(g)) in Figure \ref{fig:fig5}. The quasi-periodic intensity change can be best discerned in the red rectangle region as shown in Figure \ref{fig:fig5} from 01:38 to 01:56 UT. The intensity change underwent about 4 periods during 18 minutes of ``bright-dim-bright-dim'' mode with an average period of 4.5 minutes, which is consistent with the period obtained from the stack plots in Figure \ref{fig:fig2} and the period obtained from the wavelet analysis of light curve `L1'.

\subsection{Quasi-periodic intensity oscillations of the jet-like structures in flare ribbons}
\begin{figure*}[!ht]
    \centering
    \includegraphics[width=0.90\textwidth]{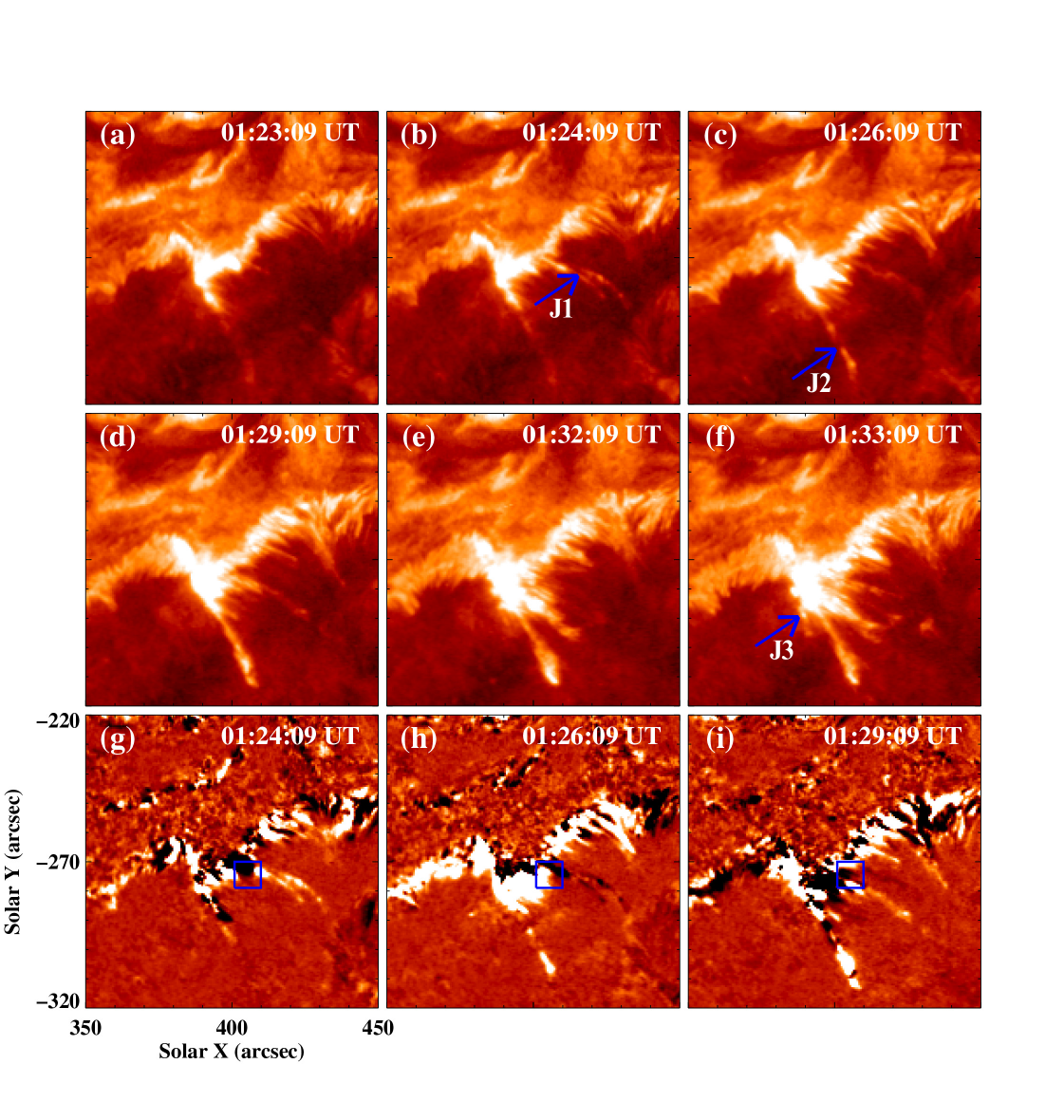}
    \caption{Panels (a)-(f): Original images at AIA 304\AA ~with the FOV shown as the white rectangle in Figure \ref{fig:fig1}(c) of a part of SR. Panels (g)-(i): Running difference images at 304\AA ~ with the same FOV as panels (a)-(f). Blue squares in panels (g)-(i) are used to obtain the intensity profile in Figure \ref{fig:fig7}(a).}
    \label{fig:fig6}
\end{figure*}

\begin{figure*}[!ht]
    \centering
    \includegraphics[width=0.90\textwidth]{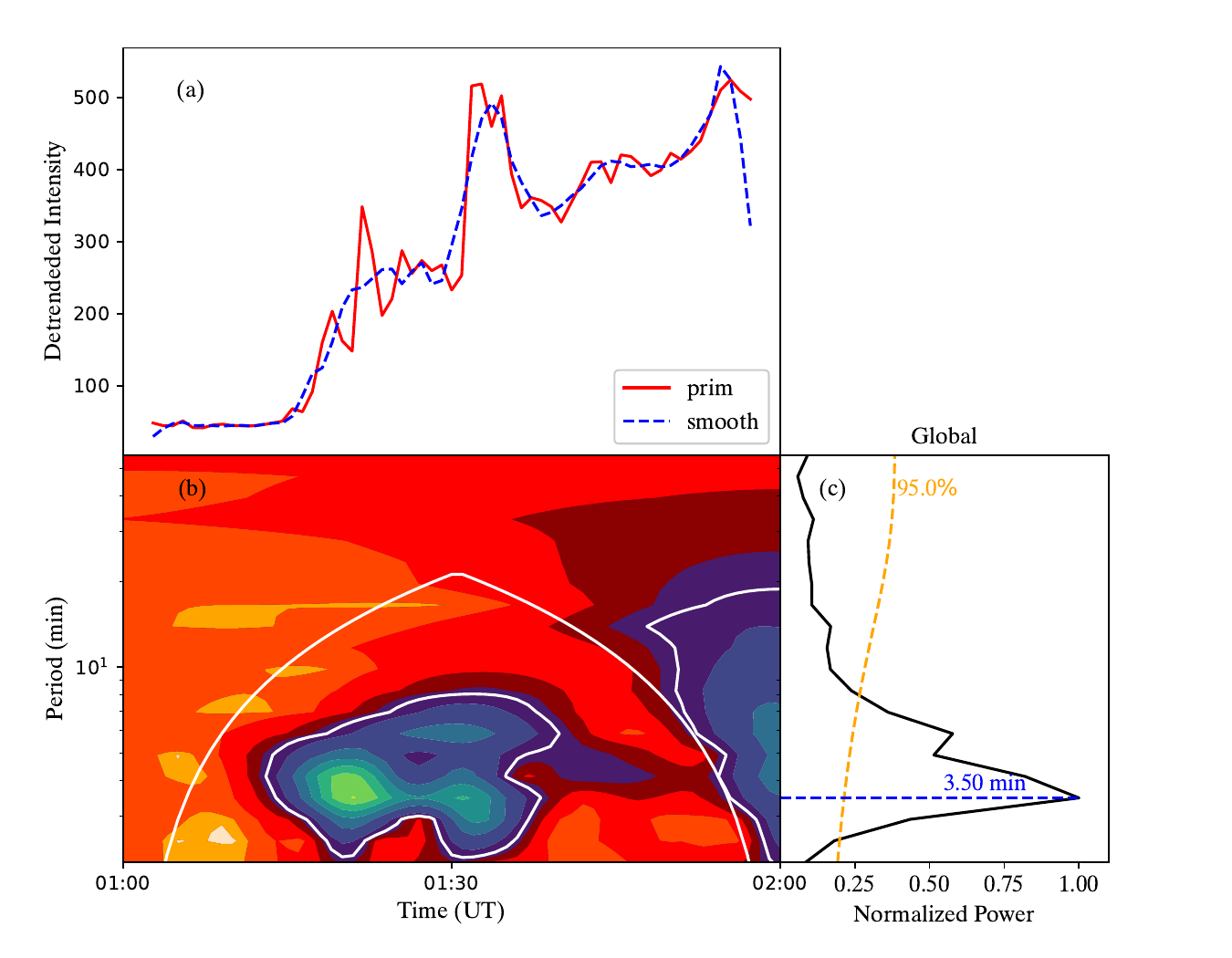}
    \caption{Panel (a): The normalized intensity integrated in the blue square at 304\AA~ shown in Figures \ref{fig:fig6}(g)-(i) (red solid line) and the smoothed light curve (blue dashed line), respectively. Panel (b): The wavelet power map of the detrended intensity from 01:00 to 02:00UT. Panel (c): The corresponding global power. The orange dashed line represents the 95\% confidence level, and the blue dashed line denotes the period over the 95\% confidence level with the max power (3.50 minutes).}
    \label{fig:fig7}
\end{figure*}

From the observations by SDO/AIA at 193 and 304 \AA,~we can see numerous jet-like structures rooted at the flare ribbons SR and NR2. The jet-like structures first appear at about 01:10UT, just before the onset of the flare. The development of the jet-like structures also lasted for a long time. From Figures \ref{fig:fig6}(a)-(f), three jet-like structures (indicated by the blue arrows `J1', `J2' \& `J3') moved at speeds of about 60-70 km/s. The footpoints of the jets in SR also displays a `bright-dim' intensity oscillation (see Figures \ref{fig:fig6}(g)-(i)). The intensity oscillation behaves similarly to that at 94\AA~in Figure \ref{fig:fig5}. To calculate the oscillation period, we perform a wavelet analysis of the total intensity restricted in the blue square. The Morlet spectrum is given in Figure \ref{fig:fig7}, and the calculated period is 3.5 minutes which is over the 95\% confidence level. We also checked the jet-like structures rooted in NR2. They behave similarly to that rooted in SR. By performing wavelet analysis of the jet-like structures in NR2, we obtain a period of 4.13 minutes.

The formation mechanism of jet-like features located in flare ribbons remains elusive. \citet{Li2019} first reported 15 jet-like features happening during an X-class flare. They suggested that these features are probably results of chromospheric evaporation. According to their study, after the onset of the two-ribbon flare, the thermal and kinetic energies released by the magnetic reconnection process will be transported through tenuous corona to the cooler and dense chromosphere. When the energy input to the chromosphere exceeds what can be shed by radiative and conductive losses, chromospheric material is heated up to a temperature on the order of 10 MK. The overpressure of the chromosphere will initiate the evaporation. In our study, we suggest that the jet-like structures in this flare may also be results of chromospheric evaporation. As the flare loops behave a quasi-periodic intensity oscillation, the material from the flare loops to the ribbons will modulate the chromospheric evaporation.

\subsection{Slipping motion at the tip of flare ribbons}
\begin{figure*}[!ht]
    \centering
    \includegraphics[width=0.90\textwidth]{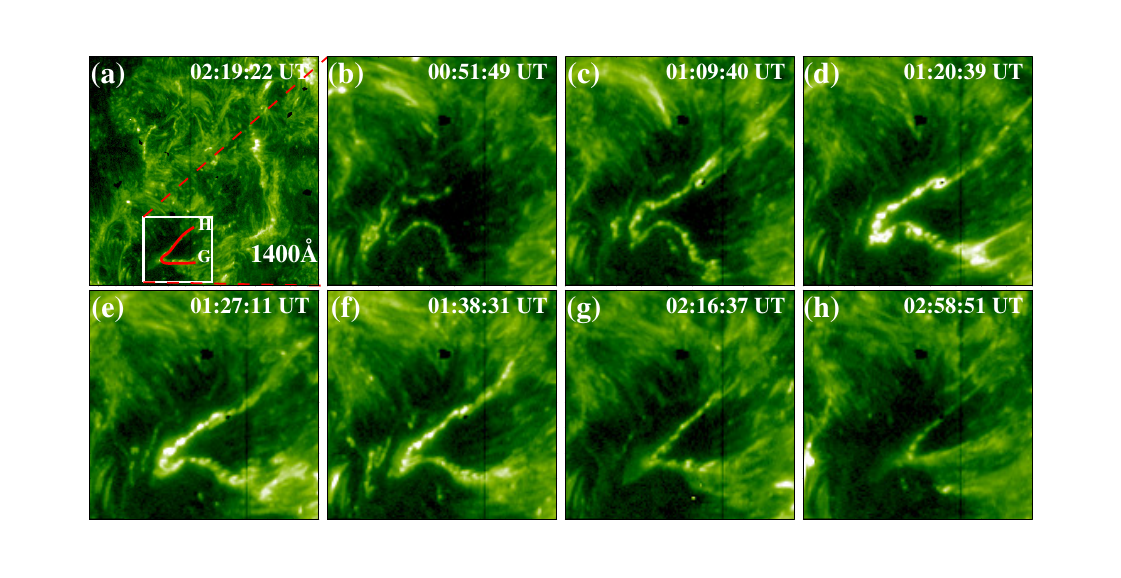}
    \caption{IRIS 1400\AA ~images of the flare ribbon NR1 with the FOV of panel (a) of 105"$\times$ 105". The FOV of panels (b)-(h) is shown in Figure \ref{fig:fig1}(b) and panel (a) as the white rectangle. The red curve in panel (a) with `G-H' denotes the slit to obtain the stack plot in Figure \ref{fig:fig9}(a). The animation of this figure includes IRIS 1400\AA~images from 00:46 to 03:05UT. The video duration is 40s.}
    \label{fig:fig8}
\end{figure*}

\begin{figure*}[!ht]
    \centering
    \includegraphics[width=0.90\textwidth]{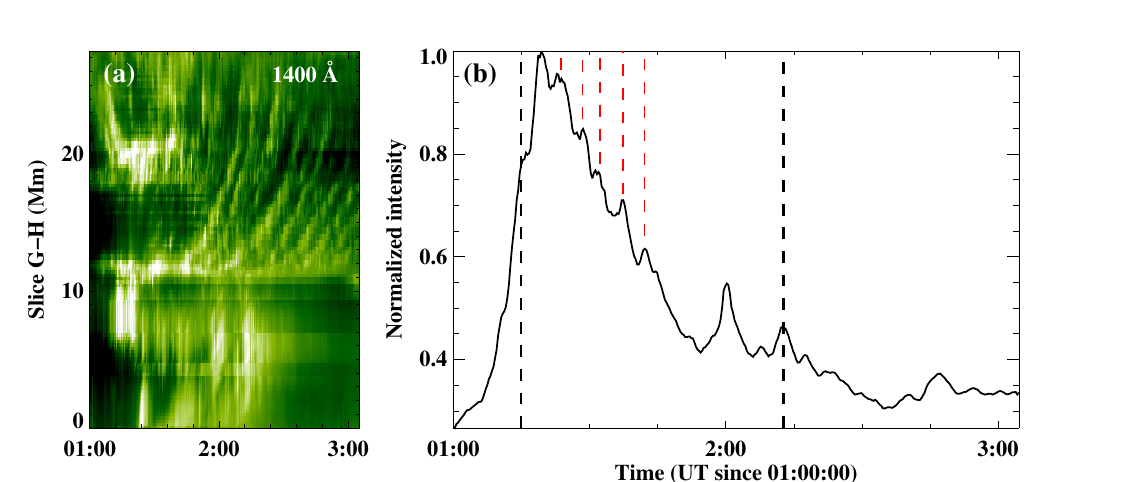}
    \caption{Panel (a): Stack plot along slice `G-H' (red solid line in Figure \ref{fig:fig8}(a)) in IRIS 1400\AA ~images. Panel (b): Black solid curve indicates the integrated intensity of 1400\AA~within the region of Figures \ref{fig:fig8}(b)-(h). Black vertical dashed lines indicate the start time and peak time of the flare, respectively. Red dashed lines highlight the oscillation peaks in the light curve.}
    \label{fig:fig9}
\end{figure*}

During the flare evolution, the footpoints of the flare loops exhibit apparent slipping motion along the flare ribbon NR1 (as illustrated in Figure \ref{fig:fig8}). The bright dot-like substructures in NR1 move from the southwestern side to the northwestern side (through the zip shape of the ribbon). Before the onset of the C9.1 flare, bright dots in the flare ribbon NR1 can be clearly seen in IRIS 1400\AA ~images (Figures \ref{fig:fig8}(b)-(c)). With the evolution of the flare, the bright knots became brighter and the slipping motion started (Figures \ref{fig:fig8}(d)-(g)) during the whole flare duration (see animation at 1400\AA). The fine structures in the ribbon became dimmer after the flare ended (see Figure \ref{fig:fig8}(h)). For a more direct display of this slipping motion, we have constructed
the stack plot along slice `G-H', visualized in Figure \ref{fig:fig9}(a). The stack plot shows multiple moving stripes in 1400\AA, each of which shows the slipping fine structure in slice `G-H'. The bright dots in the flare ribbon can propagate over 10 Mm in about 20 minutes. At least seven stripes can be identified from Figure \ref{fig:fig9}(a) during the flare. The temporal evolution of the integrated intensity within the region of Figures \ref{fig:fig8}(b)-(h) is elaborated by the black solid line in Figure \ref{fig:fig9}(b). The average time interval between neighboring stripes is calculated to be about 4.6 minutes.

\begin{figure*}[!ht]
    \centering
    \includegraphics[width=0.90\textwidth]{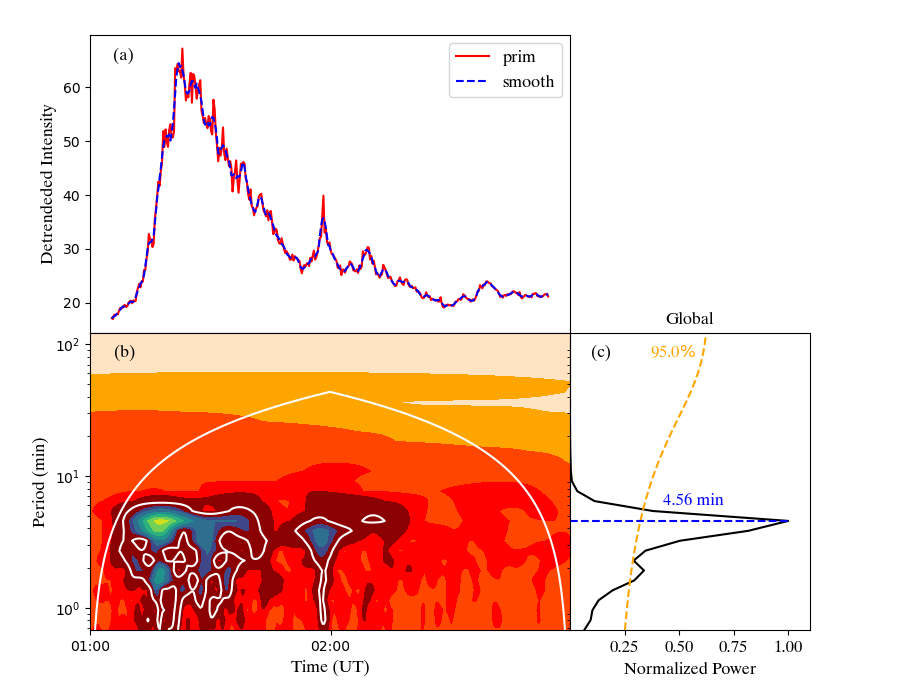}
    \caption{Panel (a): Red solid and blue dashed lines indicate the original and smoothed light curves of that in Figure \ref{fig:fig9}(b) at IRIS 1400\AA. Panel (b): The wavelet map of the detrended intensity calculated from the difference between the two light curves in panel (a) from 01:00 to 03:00 UT. Panel (c): The corresponding global power. The orange dashed line represents the 95\% confidence level, and the blue dashed line denotes the period over the 95\% confidence level with the max power (4.56 minutes).}
    \label{fig:fig10}
\end{figure*}

We perform the wavelet analysis of the light curve at IRIS 1400 \AA ~in Figure \ref{fig:fig10}. The obtained period of the ribbon oscillations is 4.56 minutes. This period corroborates well with what we calculate from the stack plots along NR1. In addition, this period matches well with that of the stationary intensity oscillations of flare loops. Based on the similar periodicity between flare ribbons and flare loops, we conclude that these quasi-periodic behaviors may share the same physical nature.

\section{Discussion}  \label{sec:discussion}
In our study, we report persistent (over 3 hours) stationary intensity oscillations of the flare loops, and they manifest as a quasi-periodic pattern with a period of 4-5 minutes at EUV bandpasses during a C9.1 flare. The long-duration stationary intensity oscillations in the flare loops, the slipping motion at the tip of the flare ribbons and also the jet-like structures rooted in the flare ribbons show the similar periodicity, which is first reported to our knowledge. To investigate what mechanism drives this periodicity, we discuss several potential possibilities.

In our observations, the period given by the EUV bandpasses is between 4 and 5 minutes. Considering only the similar time scale, the leakage of sunspot oscillations \citep{Yuan2011,Yuan2016,Yuan2023,Wang2024} could be a possible driving mechanism. However, only ribbon NR1 is located in the sunspot region (Figure \ref{fig:fig1}(a)), while the flare loops with a rather large range (over 90Mm) share stationary intensity oscillations (see Figure \ref{fig:fig2} of the stack plots over the whole flare ribbons). Thus, based on the magnetic configuration of this flare, we tend to believe that this long-duration of stationary intensity oscillations may not come from the leakage of sunspot oscillations.

For MHD oscillations in the coronal environment, fast-mode waves such as kink mode and sausage mode oscillations are commonly considered as the driving source of QPPs. The quasi-periodic behavior modulated by kink oscillations exhibits transverse displacements along the loop-like structures \citep{Li2023,Zhong2023}. However, this transverse displacement along the flare loops could not be recognized from our event. According to \citet{Lidong2023}, there are two kinds of kink oscillations. The ``decaying oscillations'' have a quite large amplitude, which is larger than the pixel size of SDO/AIA($\sim$0.4 Mm). As for the other kind of kink oscillation, ``decayless oscillation'', its amplitude is about 0.2 Mm which is smaller than the pixel size of AIA \citep{Anfinogentov2013}. Although the decayless kink oscillation is hard to observe directly from the animation due to its small amplitude, it could still be recognized from the time-distance plots. \citet{Anfinogentov2015} found the average amplitude of the decayless oscillations being 0.17 Mm among 21 active regions that could still be determined at the sub-pixel level. \citet{Mandal2022} also concluded the amplitudes of the kink oscillations being 0.1-0.5 Mm by the use of EUI on board Solar Orbiter and SDO/AIA. So based on the previous studies and instruments, the absence of transverse displacement in the time-distance plots indicates that the kink oscillations should be ruled out. Sausage modes are commonly used to explain QPPs with periods of seconds or tens of seconds \citep{McLaughlin2018, Libo2020, Zimovets2021}. \citet{Tian2016} studied QPPs in the flare loops during an M-class flare with a period of $\sim$25s. By calculating the associated phase velocity of the waves, they suggest that these QPPs are caused by the global sausage modes, since the phase velocity of the global sausage oscillations is smaller and close to the external Alfv\'en speed \citep{Nakariakov2003,Antolin2013}. The time scale of the sausage mode is similar to the transverse Alfv\'en time scale \citep{Guo2016}. The obtained period of our observation is inconsistent with the time scale of the sausage mode.

In conclusion, we tend to ascribe the long-duration stationary intensity oscillations of the flare loops in our study to the interaction of the reconnection outflow and the reconnected flare loops. Some previous studies have discussed how the reconnection downflow leads to QPP patterns. \citet{Hayes2019} studied an X-class flare event and found that the pulsation has a period of $\sim$65s during the impulsive phase and $\sim$150s during the decay phase. They believe that the persistent QPP in the decay phase is related to the interaction between the flare loops and the downward plasmoids at the loop top. \citet{Reeves2020} studied the Doppler velocity oscillations in the Fe XXI line with a period of $\sim$400s. They suggest that these oscillations are related to the reconnection outflows disturbing the loop top region. For numerical studies, \citet{Takasao2016} discovered the local oscillation at the loop top based on 2D MHD simulations, which is a result of the backflow of the reconnection outflow with a period of a few 10s to a few 100s, depending on the magnetic field strength. \citet{Shibata2023} performed 3D MHD simulations and found that the loop top region is filled with turbulent flows and shows an oscillation in Doppler velocity driven by the backflow of the reconnection outflow. In our observation, we find that the intensity oscillations of the flare loops and the slipping motion of the flare ribbons share a common period of 4-5 minutes, similar to the periods in previous studies \citep{Takasao2016,Reeves2020}. Based on previous simulations \citep{Takasao2015,Takasao2016,Takahashi2017,Ye2020,Shibata2023}, we speculate that the oscillations observed in our event may be related to the magnetic tuning fork effect during the flare process. The backflow of the reconnection outflow compresses the gas and magnetic field around the arms of the magnetic tuning fork, which drives the oscillation at the loop top region. The oscillation in the loop top region probably further results in the intensity oscillations of the flare loops. At the ribbon tip region, due to the continuous magnetic reconnection process of the magnetic field lines, the flare ribbon exhibits the slipping motion quasi-periodically. In our study framework, we claim that the quasi-periodic stationary intensity oscillations of the flare loops and the quasi-periodic slipping motion along the flare ribbons can be regarded as various aspects of the same physical process.

\section{Summary and Conclusions}  \label{sec:summary}

In this work, we report a unique case with both the long-duration stationary intensity oscillations of flare loops and the apparent quasi-periodic slipping motion of flare ribbons during a C9.1-class flare on 2015 March 15. It shows a period of 4-5 minutes in flare loops and ribbons, which is possibly related to the loop-top dynamics caused by the reconnection outflow impinging on the flare loops. The slipping motion within the flare ribbons offers a strong piece of evidence for 3D slipping magnetic reconnection. As indicated by previous theories of 3D reconnection, the footprints of magnetic lines slip along the QSL, showing a series of bright dots slipping along the flare ribbons. This corresponds well to the slipping motion of the eruptive flares observed by IRIS and SDO. However, different from previous observations \citep{Dudik2014,Li2015}, we first show both the long-duration (over 3 hours) stationary periodic intensity changes of high-temperature flare loops rooted in the straight part of the ribbons along with the quasi-periodic slipping motion at the tips of ribbons in one event. The period of the stationary intensity oscillation in the flare loop is consistent with the slipping period in the flare ribbon, which implies that the similar physical mechanism between the stationary intensity change and the slipping motion. Under the weak magnetic guide-field circumstances, the flare loops rooted in the straight part of the flare ribbon display the stationary intensity oscillations. This is similar to 2D or 2.5D models, which can be explained by 2.5D simulations. However, the quasi-periodic slipping motion occurring in the tips of the flare ribbon implies the occurrence of 3D magnetic reconnection and should be considered in 3D frameworks. We note that this configuration accords well with the 3D standard flare model developed by \citet{Janvier2014}. In the future, we will carry out 3D numerical simulations to analyze the formation mechanisms of the periodicity shown in the flares.

\section*{Acknowledgments}
We acknowledge the SDO/AIA and HMI for providing data. IRIS is a NASA small explorer mission developed and operated by LMSAL with mission operations executed at NASA Ames Research center and major contributions to downlink communications funded by the Norwegian Space Center (NSC, Norway) through an ESA PRODEX contract. This work is supported by the National Key R\&D Program of China (2019YFA0405000, 2022YFF0503800, 2021YFA1600500), the B-type Strategic Priority Program of the Chinese Academy of Sciences (XDB0560000,XDB41000000), the National Natural Science Foundations of China (12222306, 12273060, 12073073 and 11933009), and grant 202101AT070018 associated with the Applied Basic Research of Yunnan Province. Y.J. acknowledges the support by grants associated with the Yunnan Revitalization Talent Support Program, the Foundation of the Chinese Academy of Sciences (Light of West China Program).

\bibliography{sample631}{}
\bibliographystyle{aasjournal}
\end{document}